\documentclass[fleqn,usenatbib]{mnras}

\usepackage[T1]{fontenc}
\usepackage{ae,aecompl}

\usepackage{graphicx}	
\usepackage{amsmath}	
\usepackage{amssymb}	
\usepackage{tablefootnote}
\usepackage{threeparttable}

\title[HST/WFPC2 imaging analysis and modelling of MSPNe]{HST/WFPC2 imaging analysis and Cloudy modelling of the multiple shell planetary nebulae NGC 3242, NGC 6826 and NGC 7662}

\author[D. Barr\'{\i}a]{
D. Barr\'{\i}a,$^{1}$\thanks{E-mail: daniela.barria@ucn.cl}
and S. Kimeswenger,$^{1,2}$
\\
$^{1}$Instituto de Astronom\'ia, Universidad Cat\'olica del Norte, Avenida Angamos 0610, Antofagasta, Chile\\
$^{2}$Institut f\"{u}r Astro- und Teilchenphysik, Leopold Franzens Universit\"{a}t Innsbruck, Technikerstrasse 25, 6020 Innsbruck, Austria
}
\date{Accepted 2018 July 03. Received 2018 June 29; in original form 2018 March 20.}

\pubyear{2018}

\begin{document}
\label{firstpage}
\pagerange{\pageref{firstpage}--\pageref{lastpage}}
\maketitle

\begin{abstract}
We performed a detailed photometric analysis and photoionisation modelling on three high excitation multiple shell planetary nebulae: NGC 3242, NGC 6826 and NGC 7662.
Archival HST/WFPC2 narrow band filter images were used to investigate shocked regions by two independent methods: using low excitation ions (via H$\alpha$/[\ion{N}{II}] vs H$\alpha$/[\ion{S}{II}] extended diagnostic diagrams) and by means of high excitation species looking for regions of enhanced [\ion{O}{III}]/H$\alpha$ line ratios. Shocked region analysis via low excitation ions shows that major deviations from the general inside to outside ionisation trend correspond only to regions where FLIERs or LISs are located. The reduction on the SNR at the outskirts of the [\ion{O}{III}]/H$\alpha$ ratio maps, made us unable to unambiguously identified for an enhancement on the [\ion{O}{III}]/H$\alpha$ line ratio as an indicator for shocks. For non-shocked regions we performed a photoionisation modelling using Cloudy. Fittings to the [\ion{O}{III}] and H$\alpha$ observed radial profiles lead us to constrain on the free parameters of the density laws and filling factors together to temperatures and luminosities for the CSPNe.
 Best fit models show a very well representation of the [\ion{O}{III}] and H$\alpha$ emission. Discrepancies in the model fittings to the [\ion{N}{II}] and [\ion{S}{II}] profiles at NGC 6826 and NGC 3242, can be attributed in the former case, due to a contamination by the light of the CSPN and, in the latter case, either due to gas inhomogeneities within the clumps or to a leaking of UV radiation.
\\
\end{abstract}



\section{Introduction}
The standard Interacting Stellar Winds (ISW) model on the origin of Planetary Nebula (PN singular, PNe plural) establish that PNe are formed due to the interaction of a fast and tenuous wind developed at the post-AGB phase, which compress and accelerates the material ejected during the AGB lifetime \citep{kwok78}.
While the ISW model can describe the formation of macro structures such as the shell and halo in round/elliptical PNe \citep[see e.g.][]{schonberner02, perinotto04a, schonberner16}, is unable to explain the observed additional micro structures like filaments, knots, clumpiness and/or collimated outflows \citep[see e.g.][]{matsuura09,ottl14,fang15}. Certainly these structures have to be the result of radiative and dynamical processes taking place in the interaction of the stellar winds \citep{garciasegura06}, and thus make evident that the mechanism(s) involve in the nebulae formation can not be that simple as ISW model predicts. Even more, clumpiness seems to play a significant role in the ionisation structure of the gas, as is shown in a recent investigation by \citet{ottl14} on the PN NGC 2438. By means of photoionisation radiative transfer models using Cloudy\footnote{www.nublado.org}, authors conclude that the excitation and temperature throughout the nebula, and beyond, requires small-scale clumps to obtain a self consistent model.\\
Additional to the observed micro structures, extra outer shells and/or halos have been found at some PNe \citep[see e.g.][]{gomezmuñoz15,guerrero99,corradi07}. The origin of these morphologies remain uncertain but most likely they are relics of different episodes of mass loss at the last stages in the AGB phase. Considering a frequency of occurrence estimated to be around 25\%-60\% \citep{chu87,stanghellini97}, these so called Multiple Shell Planetary Nebulae (MSPNe) are thus perfect candidates to investigate macro/micro structures at PNe as well as the previous mass loss history of the post-AGB star and CSPN.\\
On the other hand, it is known that PNe can be affected by shock fronts which can move through the medium disturbing the energy balance of the gas \citep{meaburn91,dyson92,arthur94}. Previous hydrodynamic models show that shock fronts might play a role particularly at the boundary layers \citep{corradi03,corradi07,perinotto04a}.
Some investigations use the line ratios $\log$(H$\alpha$/[\ion{N}{II}]) versus $\log$(H$\alpha$/[\ion{S}{II}]) to identify photoionised PNe from shocked nebulae, such as found in supernova remnants or in \ion{H}{II} regions. This scheme was introduced by \citet{garcialario91}, refined in \citet{magrini03} and extensively reviewed by \citet{frew10}. An extension of this diagnostic diagram to two dimensions was presented in \citet{ottl14}.\\
More recently, \citet{guerrero13} showed that in regions with higher temperature, the [\ion{O}{III}]/H$\alpha$ ratio is a better indicator for shock fronts, as the low ionisation species [\ion{N}{II}] and [\ion{S}{II}] are not sufficiently abundant. This is somewhat similar to the use of [\ion{O}{III}]/H$\beta$ ratio in the classic BPT \citep{BTP81} diagram used to classify emission-line spectra from extra-galactic objects. \citet{frew10} showed that near all galactic PNe of their large sample appear superimposed to Seyfert 2 galaxies in their BPT diagrams.\\
However, as \citet{guerrero13} mentioned, the HST (Hubble Space Telescope) H$\alpha$ images obtained with the F656N filter, might be strongly contaminated by [\ion{N}{II}] emission.
Based on their analysis, the low [\ion{N}{II}] emission-line sources NGC 3242, NGC 6826 and NGC 7662 were found showing skins of enhanced [\ion{O}{III}]/H$\alpha$ line ratio, caused by a sharp increase in the local temperature at the outer edge of the shells.\\
Looking for a better interpretation of the observed properties on MSPNe, we perform a detailed photometric analysis (using narrow band HST images) and Cloudy modelling on three high excitation MSPNe: NGC 3242, NGC 6826 and NGC 7662.\\
At the first part of this paper we present the results on shocked region analysis performed via two independent methods:
 \begin{itemize}
 	\item using low excitation ions via the (H$\alpha$/[\ion{N}{II}]) vs (H$\alpha$/[\ion{S}{II}]) line ratios, following the ionisation class along the whole nebulae and,
 	\item by means of high excitation ions, using radial profiles of the [\ion{O}{III}]/H$\alpha$ ratio maps in a similar way to \citet{guerrero13}, but making use of [\ion{N}{II}] subtracted F656N H$\alpha$ images.
 \end{itemize}
In the second part of this work, we show the results of our photoionisation Cloudy best fit models to the observed H$\alpha$, [\ion{N}{II}], [\ion{S}{II}] and [\ion{O}{III}] radial emissivity profiles at non-shocked regions.
\section{HST data analysis}
Archival HST Wide Field Planetary Camera 2 (WFPC2) images of NGC 3242, NGC 6826 and NGC 7662 were obtained from the Hubble Legacy Archive (HLA)\footnote[2]{https://hla.stsci.edu/}. From the different data sets on each nebulae, we selected only those which include observations in the narrow band F656N (H$\alpha$ + [\ion{N}{II}]), F658N ([\ion{N}{II}]), F673N ([\ion{S}{II}]) and F502N ([\ion{O}{III}]) filters at the same epoch. In this way, and according to the comments from \citet{guerrero13} related to strong contamination of [\ion{N}{II}] into the F656N filter, we were able to subtract the F658N ([\ion{N}{II}]) filter contribution from the F656N (H$\alpha$ + [\ion{N}{II}]) filter image. \\
All selected images are pre-processed HST level 2 data which have been aligned north-up, re-sampled to a uniform pixel grid, same epoch combined and astrometrically corrected. Observational details on the different selected data sets are presented in Table~\ref{tab:tab1}.
\begin{table}
	\caption{Detailed summary on used HST/WFPC2 images. All images were obtained during HST program ID 6117.}
	\label{tab:tab1}
	\centering
	\begin{tabular}{l c c c}
		\hline\hline
		Target & Filter & Exp.Time [s] & UT date \\
		\hline
		NGC 3242&F656N&100&1996-04-19\\
		        &F658N&1200&1996-04-19\\
		        & F502N &40&1996-04-19\\
		        & F673N &900&1996-04-19\\
		NGC 7662& F656N &100&1996-01-05\\
		        & F658N &1200&1996-01-05\\
           	    & F502N &100&1996-01-05\\
		        & F673N &1100&1996-01-05\\
		NGC 6826& F656N &100&1996-01-27\\
		 		& F658N &1200&1996-01-27\\
		        & F502N &100 &1996-01-27\\
		        & F673N &1100&1996-01-27\\
		\hline
	\end{tabular}
\end{table}
Cosmic rays and bad pixels subtraction was performed using the \emph{lacos} task of IRAF \footnote[3]{http://iraf.noao.edu/}. Same epoch images from different filters were aligned using field stars and the CSPN as far as available. Flux calibrated-images were obtained using the IRAF/STSDAS task \emph{imcalc} by means of the inverse sensitivity conversion factor and the ST magnitude zero point from the calibrated data.\\
As was mentioned, the bandwidth of HST H$\alpha$ F656N filter include contamination from the [\ion{N}{II}] $\lambda 6583$\AA{} and [\ion{N}{II}] $\lambda 6548$\AA{} lines. Using the transmission curves for F656N and F658N filters we estimated a contamination in the F656N band of $39\%$ for [\ion{N}{II}] $\lambda 6548$\AA{} and $2\%$ for [\ion{N}{II}] $\lambda 6583$\AA{}. Considering the known line ratio intensity R = ([\ion{N}{II}] $\lambda 6583$\AA{} / [\ion{N}{II}] $\lambda 6548$\AA{}) = 3, we calculated a total [\ion{N}{II}] contribution in the F656N bandpass of $15\%$. Typical radial velocities of the objects in the Galaxy do not change this too much within the error margins. For instance, a 100\,km\,s$^{-1}$ velocity corresponds at H$\alpha$ to only 2.19\AA{} and thus 4.38\% of the width of the filter. H$\alpha$ pure images were thus obtained by subtracting from the F656N a $15\%$ of the H658N filter contribution. Hereafter, line ratios were calculated using these H$\alpha$ pure data.\\
For each HST filter image, diametrical and radial emissivity profiles were extracted using the astronomical image processing software \emph{XVISTA}\footnote[4]{http://ganymede.nmsu.edu/holtz/xvista/} (version 7.12). Sky subtraction was performed using the \emph{sky} task in xvista. Brightness profile cuts along an arbitrary position angle were extracted by means of the \emph{cut} task. In order to cover the whole nebulae surface a total of 36 diametrical emissivity profiles with an angular separation of $5\degr$ and centred in the CSPNe were obtained for each filter. The cuts width and linear extension (in pixels) were fixed according to the spatial dimensions of each nebulae.
\section{The sample}
\begin{figure}
 	\centering
 	\includegraphics[width=\columnwidth]{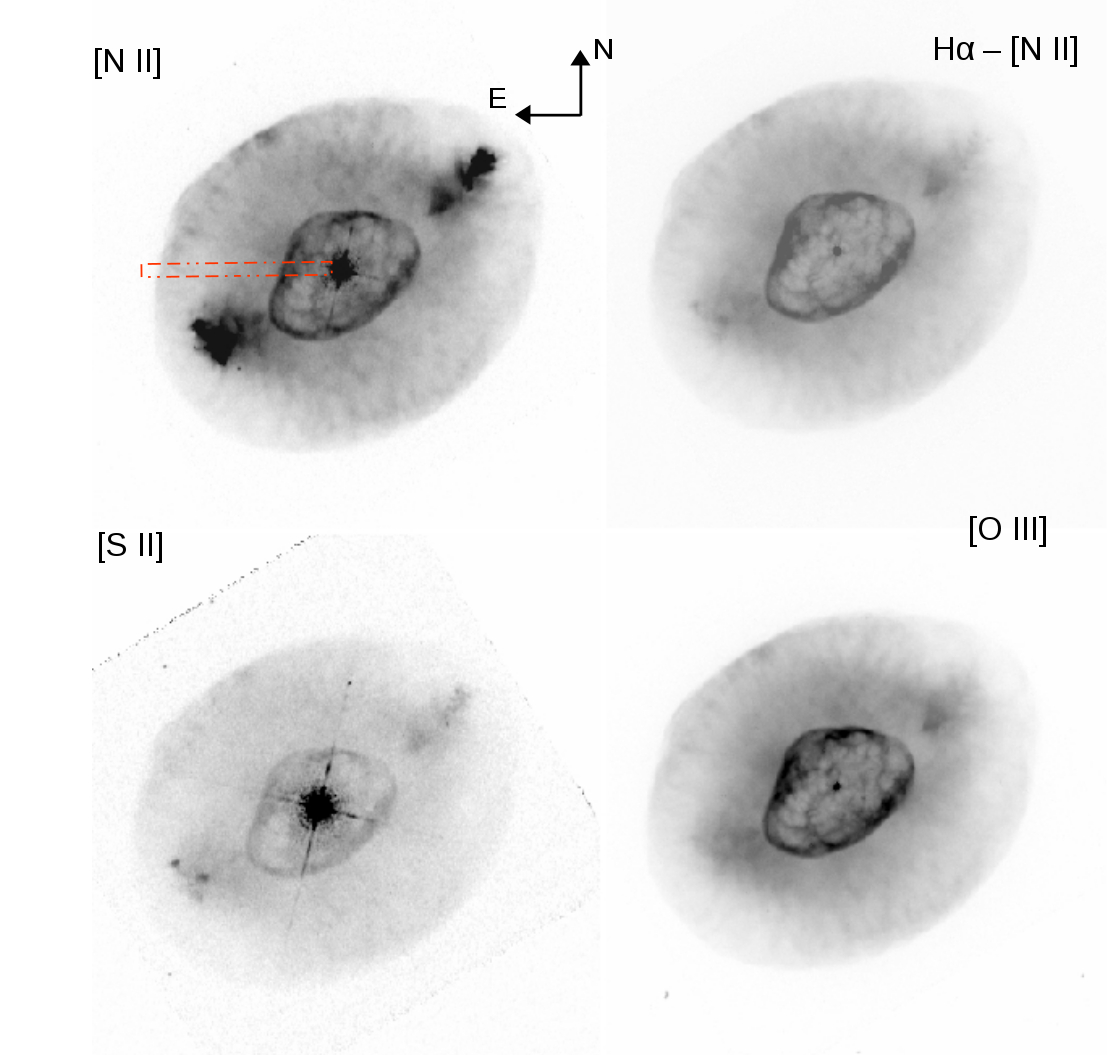}
 	\caption{\emph{Left panel:} [\ion{N}{II}] (F658N filter) and [\ion{S}{II}] (F673N filter) HST images of NGC 6826 from the HLA archive. \emph{Right panel:} Upper image shows the H$\alpha$ pure image (after [\ion{N}{II}] contribution removal), and lower image corresponds to the [\ion{O}{III}] (F502N filter). The images are displayed in an arbitrary contrast grey-scale where high intensity levels appear darker. The red rectangle indicates the region used to extract the brightness profile used for model comparison (more details see section 5).}
 	\label{fig:fig1}
 \end{figure}

\begin{figure}
  \centering
\includegraphics[width=\columnwidth]{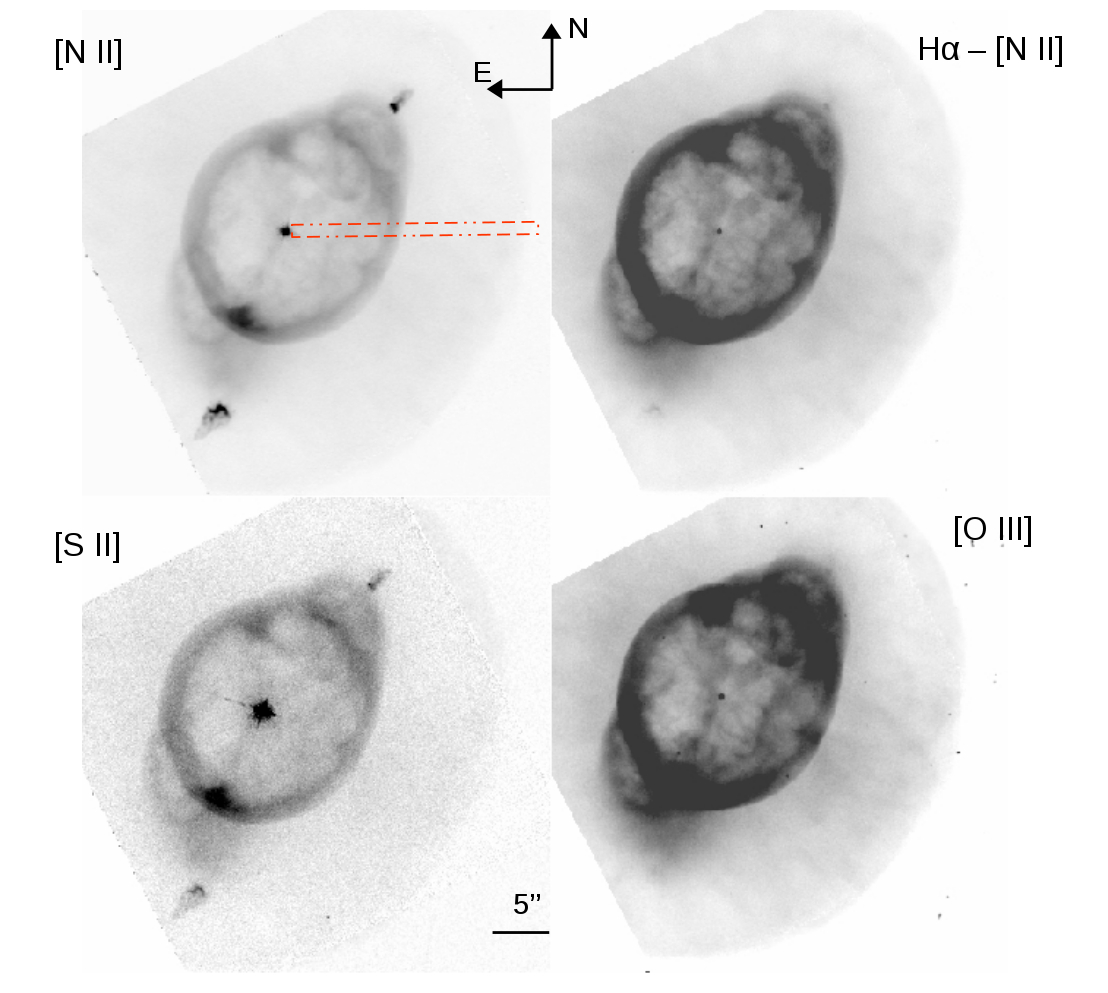}
\caption{Same as in Figure 1 but for PN NGC 3242. The two remarkable FLIERs are clearly identified in the [\ion{N}{II}] image (top left panel).}
\label{fig:fig2}
\end{figure}

\begin{figure}
	\centering
	\includegraphics[width=\columnwidth]{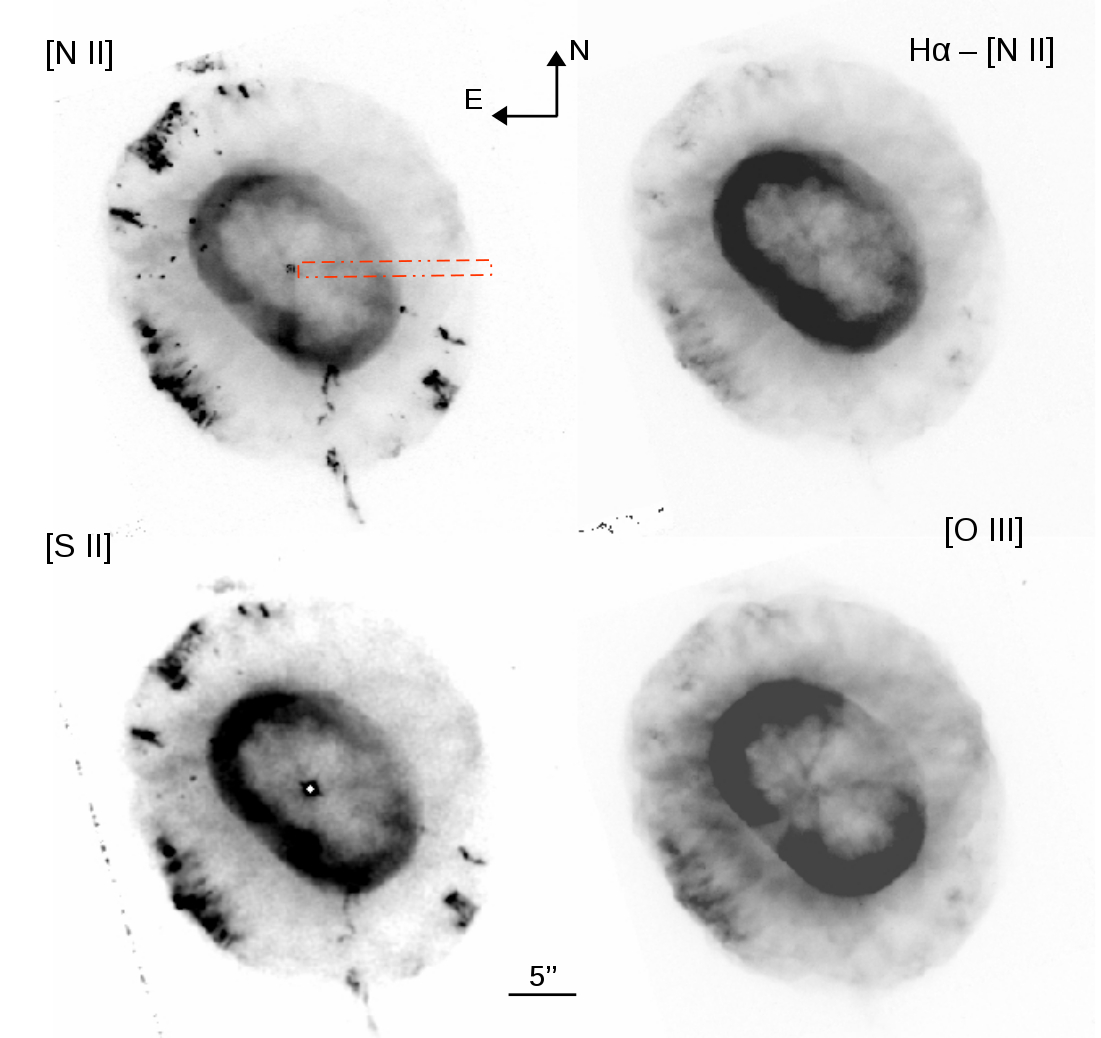}
	\caption{Same as in Figure 1 but for PN NGC 7662.}
	\label{fig:fig3}
\end{figure}

\textbf{NGC 6826:} \,This bright MSPN exhibit a 19\arcsec $\times$ 10\arcsec\, inner elliptical rim which is surrounded by an axisymmetrical outer shell of 31\arcsec $\times$ 27\arcsec. HST images show two remarkable enriched [\ion{N}{II}] Fast Low-Ionisation Emission Regions (FLIERs) which appear embedded into the outer shell (see Figure~\ref{fig:fig1}). Appreciable knots into the FLIERs support for a ionisation gradient in a similar way to NGC 3242 \citep{balick98}.
A faint and huge spherical halo extends until $\sim 130$\arcsec\,from the central star.\\
A multiwavelength spectroscopic study performed by \citet{surendiranath08} found, in general terms, that nebular abundances were similar to solar values. Exceptions were found in the case of sulphur which was found lower than solar, and carbon with a measured value a factor of two above solar. However, it become important to note than the visual spectrum used in this analysis was created by averaging several long-slit spectra taken at different location along the entire nebula. Additionally, authors used the spectral fluxes of several lines to perform a Cloudy model and thus constrain on the CSPN parameters. As input parameter they used a previously derived density profile by \citet{plait90} and no clumps were included. At an assumed distance of $\mathrm{d}=1.1\,\mathrm{kpc}$ the best fit model resulted in a central star with a temperature of $47.5\,\mathrm{kK}$, a surface gravity of $\log g=3.75$ and a luminosity of $1640\,L_{\sun}$. Previous optical spectroscopic studies performed by \citet{mendez90} resulted in a $T_\mathrm{eff}=(50\pm10)\,\mathrm{kK}$, $\log g=3.9\pm0.2$ and $\mathrm{mass}=(0.75\pm0.08)\,M_{\sun}$ of a central star. The distance was estimated by the authors in $2.2\,\mathrm{kpc}$.\\
\textbf{NGC 3242:} \,A bright and extended MSPN. It is comprised of a clumpy inner elliptical shell which is surrounded by an outer, fainter and extended elliptical shell of 42\arcsec $\times$ 39\arcsec. A pair of remarkable FLIERs oriented in the NW-SE direction appears embedded in the outer shell as is shown in the left panel of Figure~\ref{fig:fig2}. Both shells are surrounded by a fainter halo \citep{corradi03,corradi04}. By means of the expansion parallax method, \citet{gomezmuñoz15} estimated a distance to the nebula of $\mathrm{d}=660\pm100\,\mathrm{pc}$. After a morpho-kinematic study, the authors concluded that all the morphological components of NGC 3242 apparently have been originated in a clumpy medium.
\citet{miller16} found that He, C, O and Ne appear chemically homogeneous along the central cavity and the main rim of the nebula. The authors performed a Cloudy modelling using selected nebular emission lines as constrains. Unfortunately, no emission line intensity measurements (or modelling) were performed for [\ion{N}{II}] $\lambda 6548$\AA{}, $\lambda 6583$\AA{} or [\ion{S}{II}] $\lambda 6732$\AA{}. Best fit model resulted in a temperature and luminosity of $89.7^{+7.3}_{-4.7}\,\mathrm{kK}$ and $\log(\mathrm{L}/L_{\sun})=3.36^{+0.28}_{-0.22}$ for a CSPN, respectively. \\
At Figure~\ref{fig:fig2} we display the HST images of NGC 3242 used here. The main rim is fully covered on the programme 6117 images, but they do not contain the whole outer shell. At the upper right panel is shown the H$\alpha$ F656N image after removal of [\ion{N}{II}] contribution.
\\
\textbf{NGC 7662:}\, A triple-shelled early elliptical PN. The bright and apparently thick main rim extend over a 12\arcsec $\times$ 18\arcsec, surrounded by a fainter elliptical outer shell. These two main macro structures are enclosed into a faint, extended and hot circular halo \citep{chu87,sandin06}. Remarkable at NGC 7662 are the prominent low ionisation emission line structures (LISs) presented in form of several knots and a jet-like feature (see left panel in Figure~\ref{fig:fig3}). A distance of $\mathrm{d}=1.19\,\mathrm{pc}$ was determined by \citet{mellema04} taking into account correction velocity factors for plasma discontinuities (like shocks and ionisation fronts). We adopted this value for our investigation.\\
Employing HST/STIS spectra and by means of Cloudy modelling, \citet{henry15} found that best model parameters for NGC 7662 resulted in a central star of temperature $\mathrm{95\,kK}$, a luminosity of $\mathrm{2630\,L_{\sun}}$ and a mass of $0.57\,M_{\sun}$ (initial mass at the end of the AGB phase). The He/H abundance was found above solar. The authors also found remarkable discrepancies between the observed and model predictions for [\ion{S}{II}] $\lambda 6716$\AA{}, $\lambda 6731$\AA{} line strengths. To explain this, the authors proposed that either the [\ion{S}{II}] production extend beyond their imposed truncated radial distance, or shocked gas is responsible for enhancing the strength of [\ion{S}{II}]. They also mentioned for some limitations on their models, such as the use of a blackbody model for the spectral energy distribution, the assumption of radially constant gas density and the fact that they based each model on spatially integrated long-slit spectrum. Also, the filling factor was treated constantly along the nebula ($\epsilon=1.0$).
\section{Shocked regions analysis}
\subsection{2D extended diagnostic diagrams}
\begin{figure}
 	\centering
 	\includegraphics[width=7.7cm,height=21.03cm]{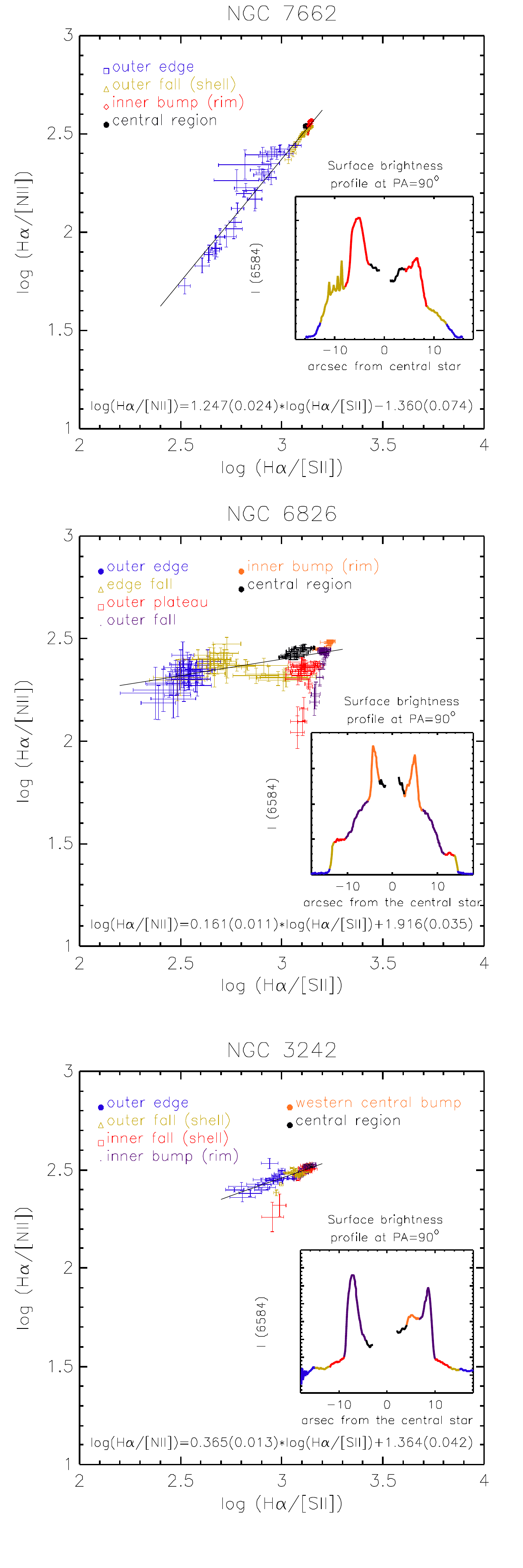}
 	\caption{Extended 2D diagnostic diagrams for NGC 7662 (upper panel), NGC 6826 (middle panel) and NGC 3242 (bottom panel). Inner plots display a extracted [\ion{N}{II}] brightness profile at $\mathrm{PA=90\degr}$ (central star contribution has been masked). Different colours corresponds to distinct regions within each nebulae, as indicated in the inserted profile pictures. Errors of fitting line parameters are given in brackets.}
 	\label{fig:fig4}
 \end{figure}

 Prior to the modelling it is necessary to scrutinised for possible shocked regions where the further photoionisation modelling will not work. [\ion{N}{II}], [\ion{S}{II}] and H$\alpha$ pure images were used to create spatial extended 2D diagnostic diagrams (hereafter E2DD) for each nebula. We looked for possible deviations at the inside to outside excitation tendency in the (H$\alpha$/[\ion{N}{II}] vs. (H$\alpha$/[\ion{S}{II}]) line ratio diagrams. To do this and in order to cover the whole nebulae surface, we extract 36 diametrical emissivity profiles with an angular separation of $5\degr$ centred in the CSPNe. The profiles were created using a rectangular box 10 pixels ($\equiv$ 1\arcsec) high and variable length (depending on the spatial extension of the nebula regions) and rotating it throughout the gas. Each of these profile was then divided by sectors according to the main fragmentation structures observed in the brightness profiles. Doing this, we were able to determine the line ratios at specific regions within the nebula. Figure~\ref{fig:fig4} shows the E2DD for each studied nebulae. Main panel illustrates the ionisation class toward the centre. Different colours indicate distinct regions within the nebulae. As an illustrative example, the inner panels display one of the used [\ion{N}{II}] brightness profile extracted at a position angle (PA) of $\mathrm{90\degr}$ with the corresponding regions marked by colours.\\
 A first inspection at Figure~\ref{fig:fig4} reveals that major deviations from the inside to outside ionisation tendency, might seems to come from regions where FLIERs or LISs are located.\\
 At the outer regions of NGC 7662, the lower ionisation contribution from the LISs (compared to the surrounding gas), slightly deviate the line ratio values from the general tendency. A fit line to the data gives an inclination of $1.247$ and a fitting standard deviation of $0.013$. Deviating points over a 2$\sigma$ level which comes from regions where LISs are located were not considered at the fitting. NGC 7662 exhibit the steepest ionisation trend of the nebulae in our sample. At Figure~\ref{fig:figA1} we show sectional E2DD obtained as a result of the 36 extracted diametrical regions along the nebula. The influence of LISs is clearly detected along the position angles $10\degr<PA<75\degr$ and $105\degr<PA<160\degr$ in agreement with the appreciable structures in the HST images (position angles have been measured from the north top and anticlockwise).
The individual E2DDs in steps of $5\degr$ in PA (see Figure~\ref{fig:figA1}) where used to derive a general tendency. While in the regions $80\degr<PA<100\degr$ and $165\degr<PA<180\degr$ approximately similar results were obtained, the sectors with the LISs deviate from the general tendency. Thus, it is assumed that the ranges outside the LISs are dominated by photoionisation as excitation mechanism, while the directions towards the LISs are excluded due to influence of the energy balance by further processes.\\
A higher dispersion is observed in the NGC 6826 diagram. The influence of FLIERs disturbs the general tendency at regions within the shell. In a similar way to NGC 7662, Figure~\ref{fig:figA2} shows the E2DD created for NGC 6826 using each extracted diametrical brightness profile with angular $5\degr$ steps. In parallel to the HST images, FLIERs appears dominating in the diagrams between $105\degr<PA<135\degr$. After a linear fitting to the data, we found a $0.161$ inclination factor with a $0.015$ fitting dispersion value.\\
Similar to NGC 6826, but with a smaller spatial extension, the presence of FLIERs in NGC 3242 also is revealed in its E2DD. As can be seen in the individual diagrams in Figure~\ref{fig:figA3}, its location in the HST images ($135\degr<PA<165\degr$) match to the deviating regions in the E2DD. Resulting best fit line to the data shows a inclination of $0.365$ with the lowest fitting standard deviation of $0.006$.\\
We conclude that no additional deviating data points beyond the ones related to FLIERs or LISs existence, has been detected in the NGC 6826, NGC 3242 and NGC 7662 E2DDs.
\subsection{[O\,III]/H$\alpha$ line ratios}
\begin{figure}
	\centering
	\includegraphics[width=\columnwidth]{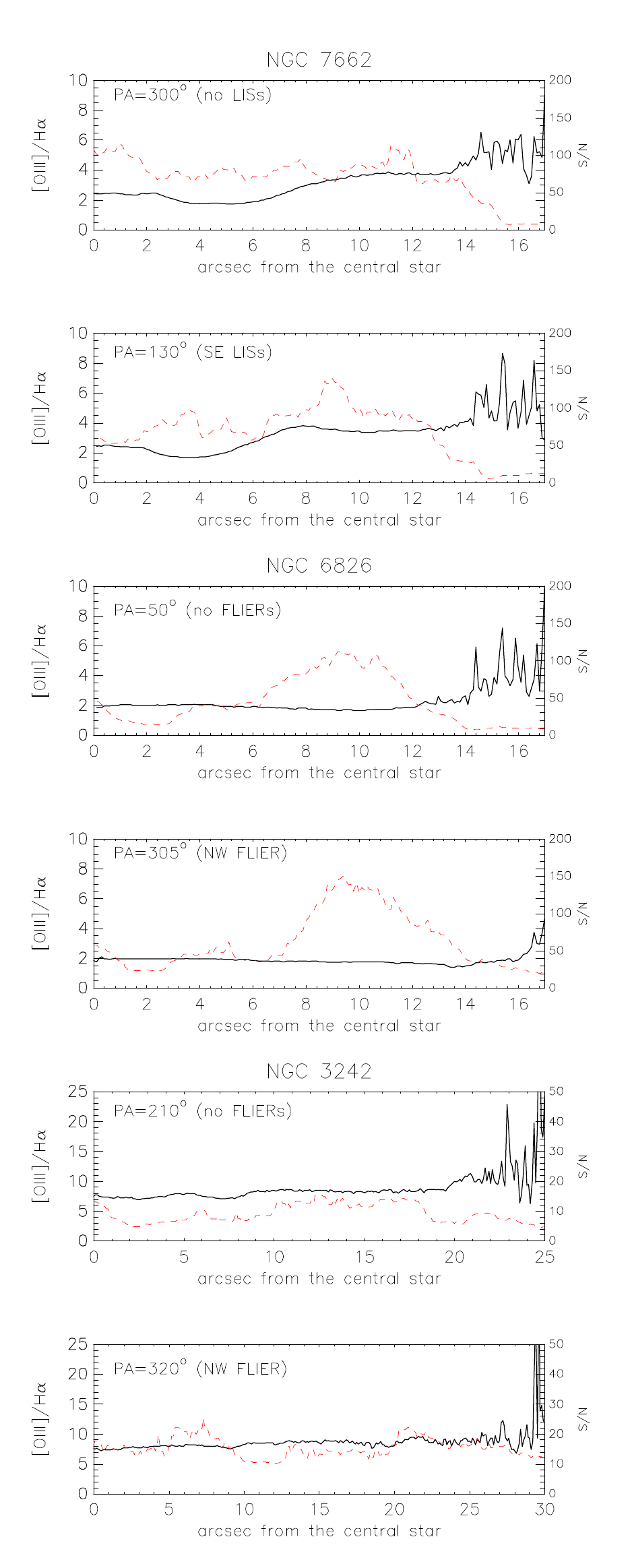}
	\caption{Selected extracted radial profiles from the [\ion{O}{III}]/H$\alpha$ ratio maps of NGC 7662 (upper panels), NGC 6826 (middle panels) and NGC 3242 (bottom panels). Solid lines corresponds to the [\ion{O}{III}]/H$\alpha$ ratio intensity and dashed line represents the SNR of the function per pixel.}
	\label{fig:fig5}
\end{figure}

\citet{guerrero13} suggest that a fast and dense shell which expands into a lower density medium can generate shock fronts which should increase the local electron temperature ($\mathrm{T_{e}}$). As a consequence of this, shock fronts would be responsible for a confined enhancement on the [\ion{O}{III}] emission (very sensitive to $\mathrm{T_{e}}$) whereas the low density would decrease the emission of the $\mathrm{H\alpha}$ line. Authors made use of HST (H$\alpha$+[\ion{N}{II}]) and [\ion{O}{III}] images to identify possible shocked regions looking for local enrichment in the [\ion{O}{III}]/H$\alpha$ line ratio. Ratio maps show skins of enhanced [\ion{O}{III}]/H$\alpha$ at the different shell's interfaces of NGC 3242, NGC 6826 and NGC 7662. Radial profiles were extracted from these ratio maps, revealing thin regions of enhanced [\ion{O}{III}]/H$\alpha$ ratio typically larger by a factor of two at the edge of the shells. The authors, however, did not explain in detail in the original work the orientations of the profiles and the assumed distances of the targets.
Guerrero (2017, priv. comm.) explained that they used several co-added radial profiles in different directions to improve the signal to noise ratio (hereafter SNR) in the outskirts. Nonetheless, the enhanced structures derived by \citet{guerrero13} are covering only one or two pixels in a region of a very high intensity gradient. At this sensitive zone a marginal displacement of the high gradient or a improper convolution in the handling of the resolution differences of HST images as function of wavelength (0\farcs052 and 0\farcs069, for H$\alpha$ and for [O III], respectively), might change such features when different regions are averaged. Furthermore the strong gradient of the flux by more than an order of magnitude and the known wide low surface brightness wing of the HST PSF makes small displacements strongly affecting the results. This is somehow similar to what is shown in case of scattered lights in integral field spectroscopy units (IFUs) in \cite{sandin06}. We consider that any proper average profile have to be using an extensive model of ellipticity or even a more complex outer geometry.\\
As a comparison, we used our $\mathrm{H\alpha}$ images corrected from [\ion{N}{II}] contamination and sectional emissivity profiles to search for possible increments on the [\ion{O}{III}]/H$\alpha$ ratio throughout the whole nebulae. At Figure~\ref{fig:fig5} are shown selected radial profiles extracted from our [\ion{O}{III}]/H$\alpha$ ratio maps. To be able to investigate the different structures and avoid confusion with FLIERS or LIS and considering the reasons mentioned above, we did not average profiles in different position angles.
Solid lines corresponds to the intensity of the [\ion{O}{III}]/H$\alpha$ ratios, while the red dashed lines show a robust estimation of the SNR of the function along the abscissa. The SNR values were calculated considering a theoretical error propagation equation (for uncorrelated functions) which estimate the SNR degradation (for an arbitrary pixel) due to the process of dividing two functions \citep{newberry91}. In this way, the SNR for the [\ion{O}{III}]/H$\alpha$ ratio function can be estimated by,
\begin{equation}
\left(\frac{S}{N}\right)_{[\ion{O}{III}]/H\alpha}=\left[1+\left(\frac{(S/N)_{[\ion{O}{III}]}}{(S/N)_{H\alpha}}\right)^2\right]^{-\frac{1}{2}}\left(\frac{S}{N}\right)_{[\ion{O}{III}]}
\end{equation}
where $\mathrm{(S/N)_{H\alpha}}$ and $\mathrm{(S/N)_{[\ion{O}{III}]}}$ represent the SNR of the individual $\mathrm{H\alpha}$ and $\mathrm{[\ion{O}{III}]}$ emissivity profiles, respectively. As can be seen in Figure~\ref{fig:fig5}, the reduction on the SNR at the outskirts of the studied nebulae (typically around 10 or smaller), together with the above mentioned averaging of profiles in different directions by \citet{guerrero13}, we are unable to unambiguously
reproduce their results. Although we are observing similar tendencies in the overall line ratio profiles.
\section{Cloudy modelling}
\begin{figure}
	\includegraphics[width=8.3cm]{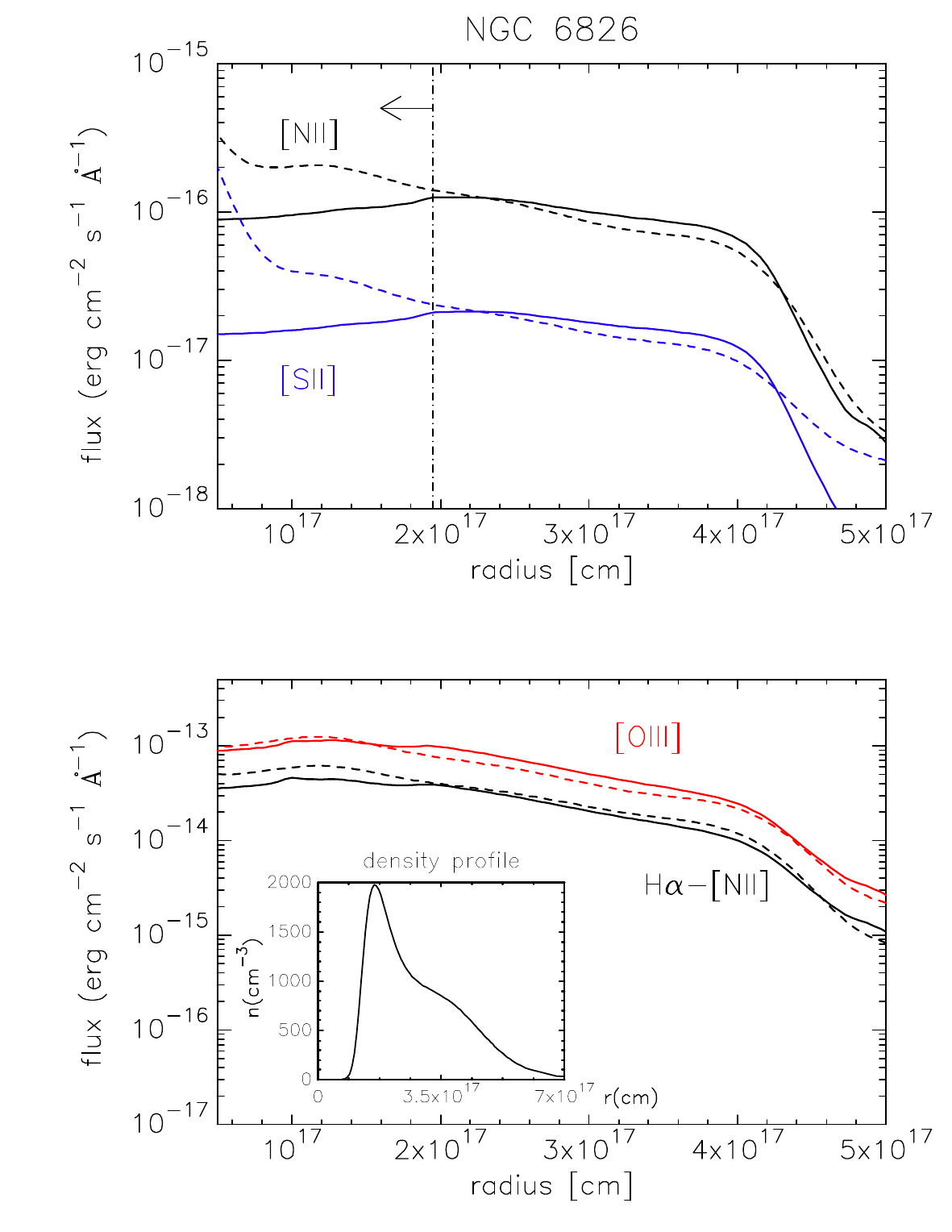}
	\caption{NGC 6826 best fit Cloudy model. The dashed lines indicates the observed [\ion{N}{II}], [\ion{S}{II}], [\ion{O}{III}] and H$\alpha$ pure emissivity profiles, while solid lines corresponds to our best fit models. The vertical dashed line and the arrow at the upper panel mark the region influencing the weak lines strongly by the stray-light of the bright central star in the images. Derived density profile is shown at the inner plot of lower panel.}
	\label{fig:fig6}
\end{figure}

\begin{figure}
	\centering
	\includegraphics[width=8.3cm]{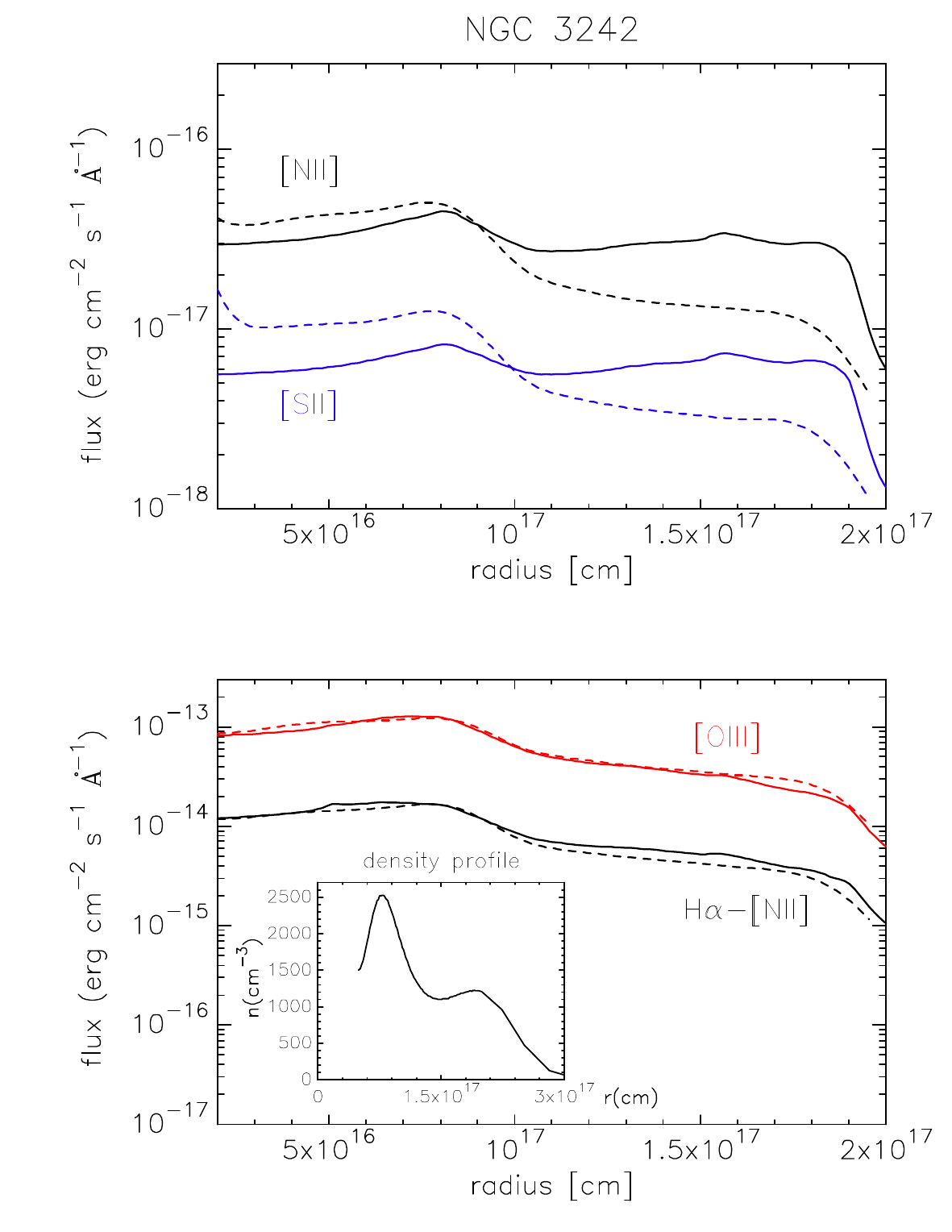}
	\caption{Same as in Figure 6 but for NGC 3242.}
	\label{fig:fig7}
\end{figure}

 Our goal is to well reproduce the observed [\ion{O}{III}], H$\alpha$ pure, [\ion{S}{II}] and [\ion{N}{II}] emissivity profiles at selected non shocked regions. According to the sectional E2DD (see Appendix A) we choose regions where the ionisation tendency from inside to outside show no major deviations and thus avoiding zones of highly variable ionisation. Red rectangles at Figures~\ref{fig:fig1}, \ref{fig:fig2} and \ref{fig:fig3} shown our selected regions to model.\\
The photoionisation modelling was performed with version 17.00 of Cloudy, last described by \citet{ferland17}.
One of the biggest advantages of using Cloudy relies on its use of extensive atomic and molecular database together to the possibility to include grains and incorporates a filling factor to simulate clumpiness.
The net line emission provided by Cloudy per volume-unit ($\mathrm{erg\,cm^{-3} s^{-1}}$), was used to create a model 2D projected emissivity profile along the line of sight by an in-house IDL code. Distance, reddening, and the expansion velocity of the outer shell ($v_\mathrm{{shell}}$) were obtained from literature (see Table~\ref{tab:tab2} for corresponding references). $v_\mathrm{{shell}}$ was used to estimate a kinematic age on each nebulae and thus constrain on the initial parameters for the central stellar effective temperature and luminosity making use of \citet{blocker95} evolutionary post-AGB models. Different initial mass (at the tip of the AGB phase) were tested. The estimated CSPNe parameters were used only as a first approach and then adjusted until reach a good match to the observations. For the spectral energy distribution of the CSPNe we used state-of-the-art H-Ni NLTE model stellar atmospheres provided by \citet{rauch03}. A static spherical geometry was assumed for all PNe and dust was treated as a free parameter thorough the analysis. The use of Cloudy option \textquotedblleft sphere expanding\textquotedblright\, had marginal impact on the final result. We used then a static solution in order to compare with similar studies like \citet{ottl14} and \citet{miller16}.\\ Instead of use a constant density or an arithmetic density law (e.g. $r^{\alpha}$; $-2<\alpha<-1$), we use non analytically given density profiles in tables. Every model was computed setting an initial H density: The observed H$\alpha$ brightness profile was used and compared with the density structure profiles from the hydrodynamic models of \citet{perinotto04a}. The data derived from the chosen density profile were then introduced as a numeric table into the model. From this starting point modifications were applied during the modelling as also used in \citet{ottl14}. Best fit density profiles are shown as inserts in Figures~\ref{fig:fig6}, \ref{fig:fig7} and \ref{fig:fig8}. Free parameters of the density laws, filling factors and central star temperatures were constrained by fitting the [\ion{O}{III}] and H$\alpha$ observed profiles to the model. The modelling process was carried out in the following terms: each nebulae was divided in two or three independent sub-shells (herafter shells (a), (b) and (c)), each of them delimited by the main observed fragmentation structures. These sub-shells
were independently modelled using filling factors as free parameters. The final model on each PN was obtained after added the resulting best models for each sub-shell. Element abundances for He, O, N and S were also adjusted during the analysis. Due to lacking images in other element species, the remaining abundances were set in Cloudy to the typical PN element abundance taken from \citet{aller83} and \citet{khromov89}.\\
At Figures~\ref{fig:fig6}, \ref{fig:fig7} and \ref{fig:fig8} are displayed the best Cloudy model fitted to the observations. Final parameters on each model are listed in Table~\ref{tab:tab2}.
\subsection*{NGC 6826}
After comparison to the theoretical model (see Figure~\ref{fig:fig6}), the observed inner regions of NGC 6826 ($7.8\times 10^{16}\,\mathrm{cm}<r_{a}<2\times 10^{17}\,\mathrm{cm}$) display an enhancement in the [\ion{N}{II}] and [\ion{S}{II}] emission.  At this zone the weaker lines appear strongly affected by the stray-light of the bright central star in the images (see upper panel at Figure~\ref{fig:fig6}). Fine tuning in the fitting procedure at this region (which covers the whole nebula rim) resulted in a $\epsilon=0.5$ filling factor.
The second sub-shell (b) was modelled spanning the full outer shell ($2\times 10^{17}\,\mathrm{cm}<r_{b}<5.7\times 10^{17}\,\mathrm{cm}$). While the predicted [\ion{N}{II}] emission fit very well to the observations, the observed [\ion{S}{II}] emission results higher at the outer edge of the shell. Considering the lower filling factor found at this region ($\epsilon=0.1$), we suggest that a geometrical effect (ellipticity) could play a significant role into the fitting process. On the other hand, observed [\ion{O}{III}] and H$\alpha$ profiles, show no major discrepancies along the nebula to our best fit model.
A comparison to previous abundance determinations \citep[see e.g.][]{liu04,barker88,surendiranath08} show no significant differences to our S abundance estimation.\\
Mostly of the previously Cloudy models used to constrain on the CSPN parameters of NGC 6826 \citep[see e.g.][]{surendiranath08,fierro11} make use of a $1\,\mathrm{kpc}$ distance to the nebula. These previous models also use either a Tlusty atmosphere model or a blackbody spectral distribution for the central star. We originally begin our modelling on NGC 6826 using a fixed $1.1\,\mathrm{kpc}$ distance and a CSPN effective temperature of $T_\mathrm{eff}=40\,000\,\mathrm{K}$ as initial assumption. After several attempts to reproduce the observations varying the CSPN temperature at the fixed distance, we were not able to reach a good match. A higher distance of $\mathrm{d}=2.1\,\mathrm{kpc}$ and CSPN temperature of $65\,000\,\mathrm{K}$ were necessary to achieve for a good fit model.
\\
Very recent results on parallaxes from the GAIA second data release \citep[therafter DR2;][]{GAIA_DR2,GAIA_DR2_PARALAXES} claim for a distance of $D=1.575^{+0.129}_{-0.111}\,\,\mathrm{kpc}$ to NGC 6826.
 This distance is $\approx$25\% shorter, but is still within the error margin of the distance of \citet{schonberner05} we used (see also Sec.~\ref{sec_con} for further discussion). The new calculated kinematic age scales linear with the distance and reduces by the same percentage (to 3300 years; $-$25\%). The evolutionary model used then predicts for small deviations from the CSPN parameters listed in Table~\ref{tab:tab2}. This is due to the fact that the [\ion{O}{III}]/H$\alpha$ ratio dominates the final value for the central star luminosity. Thus, the shorter distance will result in a somewhat lower temperature of $T_\mathrm{eff}=56\,000\,\mathrm{K}$ ($-$14\%), a marginal increase of the luminosity to $6090\,L_{\sun}$ ($+$1.5\%) and a lower surface gravity of $\log g=4.38$ ($-$0.62 dex) for a CSPN. The integrated nebular mass is the most affected parameter which then decreases approximately by 40\%.
\subsection*{NGC 3242}
Best model fits to the [\ion{O}{III}] and H$\alpha$ emission profiles of NGC 3242, displayed at the bottom panel in Figure~\ref{fig:fig7}, show almost a perfect fit to the observed emissivity. Final model have been considered after a nebular sub-division in the following regions: sub-shell (a) corresponding to the main rim of the nebula ($5\times 10^{16}\,\mathrm{cm}<r_{a}<1.0\times 10^{17}\,\mathrm{cm}$), a sub-shell (b) spanning over the main body of the shell ($1.0\times 10^{17}\,\mathrm{cm}<r_{b}<1.7\times 10^{17}\,\mathrm{cm}$) and, a sub-shell (c) covering the outer shell edge ($1.7\times 10^{17}\,\mathrm{cm}<r_{c}<2.1\times 10^{17}\,\mathrm{cm}$). Model fittings to the [\ion{N}{II}] and [\ion{S}{II}] profiles at sub-shell (a) are in good agreement to the observations, revealing a high filling factor at the inner shell where $\epsilon=0.9$. Lower filling factors were found at the outer shell ($\epsilon_{b}=0.6$ and $\epsilon_{c}=0.4$).\\
Best model result for a central star temperature and luminosity of $T_\mathrm{eff}=80\,000\,\mathrm{K}$ and $5490\,L_{\sun}$, respectively. Making use of the Stefan-Boltzmann law, we estimated a central star radius of $0.39\,R_{\sun}$. After compared stellar parameters to theoretical post-AGB evolutionary models from \citet{blocker95}, we found an initial mass (at the tip of the AGB phase) of $0.605\,M_{\sun}$ and thus a surface gravity of $\log g=5.03$. Our initial mass and central star radius are thus higher than estimations from \citet{miller16}, who predicts a $M=0.56\,M_{\sun}$ and $R=0.20\,R_{\sun}$. However, our initial mass seem to be in more agreement to the average white dwarf mass of $0.603\,M_{\sun}$ found by \citet{liebert05} and to the model result found by \citet{henry15}, who estimated a mass of $0.6\,M_{\sun}$ for the central star. \\
After compare observed nebular abundance to model abundance predictions, \citet{henry15} pointed out the large disagreement found at the S abundance.
Slit observations through a variable density region and the use of integrated light from the slit, made the authors argued for a possible misrepresentation conditions within the nebula. Our results, however, also indicates for a observed to modelled differences on the [\ion{S}{II}] emission.\\
Based on the lower filling factors found at the outer shell together to the observed/modelled emission discrepancies at the low excitation species, we propose for a leaking of UV radiation through the outer shell which leads to a much higher ionisation rate further out in the nebula. Based on this, we should expect for an overestimated emission from low excitation species at the outer regions, as can be noted at the upper panel in Figure~\ref{fig:fig7}. Additionally, it is worth noting that despite the fact that Cloudy is able to simulate clumpiness through a filling factor, the code does not take into account for possible gas inhomogeneity within the clumps themselves. The existence of cold sub-clumps embedded into bigger structures, can also might play a significant role in the emissivities profiles \citep[see][]{ottl14}.
\begin{figure}
	\centering
	\includegraphics[width=8.3cm]{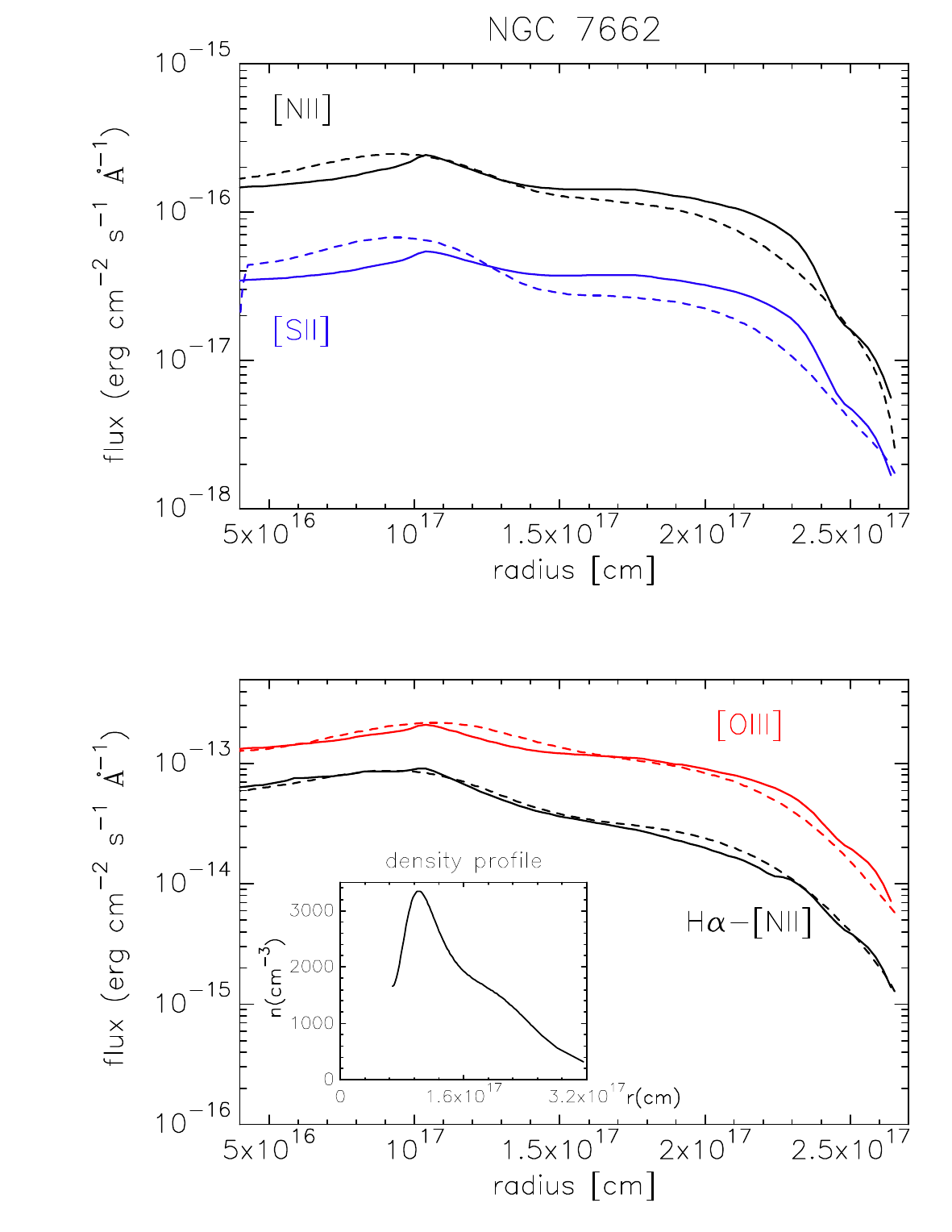}
	\caption{Same as in Figure 6 but for NGC 7662.}
	\label{fig:fig8}
\end{figure}
\\
DR2 parallaxes give a distance $D=1.466^{+0.218}_{-0.168}\,\,\mathrm{kpc}$ to NGC~3242. This distance is a factor of 2.2 times higher than that one computed by \citet{gomezmuñoz15} in their detailed spatio-kinematic study and adopted in this work. Although the GAIA statistical error is small for NGC~3242, it has a very low number of only 8 astrometric solution visibilities
({\tt visibility\_periods\_used}) compared to the 15 to 20 used for bright targets in DR2, a criterion normally used to filter DR2 catalogue quality verification \citep{arenou18}.\\
The new distance then results in a nebular kinetic age of 10\,700 years. Comparison to evolutionary models from \citet{blocker95} for the same initial mass previously used ($0.605\,M_{\sun}$), outcome in significant differences for the CSPN parameters. The temperature of the central star increases around 50\% ($T_\mathrm{eff}=124\,000\,\mathrm{K}$) as well as the surface gravity ($\log g=7.12$, +42\%). However, the most affected parameter are the central star luminosity which decreases to only $270\,L_{\sun}$ and, the nebular mass which increases even by a factor of 10. Moreover, the now very low luminosity derived from evolutionary tracks, does not represent for a bright 12\fm2 central star derived using spectroscopic distance methods \citep{heber84,napiwotzki01}.
\subsection*{NGC 7662}
Three main regions were used to build up the final Cloudy model for NGC 7662. Best fit model was reached for a $100\,000\,\mathrm{K}$ temperature and a $5250\, L_{\sun}$ luminosity of a central star. Filling factors varies from 0.3 (inner shell), 0.6 (outer shell) and 0.2 (outer edge). As can be noted in Figure~\ref{fig:fig8} no remarkable differences in the observed and modelled emissivity profiles can be attended. Our best fit model results in a $70\%$ He abundance respect to a typical galactic PN and $15\%$ above solar, whereas N and S abundances are a factor 1.5 higher than typical values.
As a comparison, \citet{zhang93} used infrared and radio archive data together to evolutionary models to reproduce a NGC 7662 central star temperature of $96\,800$ K, a luminosity of $5800\,L_{\sun}$, a surface gravity of $\log g=5.4$ and a initial mass of $0.626\,M_{\sun}$.
Density and ionisation structure of NGC 7662 were investigated by \citet{guerrero04} by means of a 3D spatio-kinematic model together with HST images. Synthetic model profiles of the H$\alpha$, \ion{O}{III}, \ion{He}{II} and \ion{N}{II} along $PA=315\degr$ (no FLIERs region) were fitted to observed surface brightness profiles at an assumed distance of $800\,\mathrm{pc}$ to the nebula. The predicted evolutionary stage of the nebula (phase II in comparison to the \cite{perinotto04a} models) for an adopted central star mass of $0.605\,M_{\sun}$, is comparable to our result. They used as best fit density profile at the inner shell a constant density $\sim 5\,000\,$cm$^{-3}$, and a power-law ($r^{-1}$) at the outer shell. Our best fit density profile (see the insert at Figure~\ref{fig:fig8}) shows a smoother density distribution in the inner region, but the outer shell appears also characterised by a function very near to a power law index with an index of -1.\\
The parallax from DR2 gives a distance of $D=1.978^{+0.343}_{-0.254}\,\,\mathrm{kpc}$ to the nebula.
However, the source was non used as primary source ({\tt astrometric\_primary\_flag=false}) for the local astrometric solution due to high astrometric noise ({\tt astrometric\_excess\_noise}).
Considering the new parallax measurement we would end up with a slightly older nebulae (5107 years) compared to our previous estimation (3080 years). Evolutionary models for a initial stellar mass of $0.615\,M_{\sun}$ predicts for a central object with a temperature of $113\,190\,\mathrm{K}$ ($\approx$13\% higher), a $2668\, L_{\sun}$ luminosity ($\approx$50\% lower) and a surface gravity of $\log g=5.96$ (+10\%). Again the integrated nebular mass increasing by a factor of 4 is the most affected parameter.
\begin{table*}
	\caption{Listed are the PNe parameters extracted from literature together to CSPNe and nebular parameters of the best fit Cloudy models.}
	\label{tab:tab2}
	    \begin{threeparttable}
	\begin{tabular}{l c c c c c c}
		\hline\hline
		 & NGC 6826   & &NGC 3242 & & NGC 7662&\\
		 Parameter & Value & Ref. & Value& Ref. & Value& Ref. \\
		\hline
	 \textit{CSPNe}&  & & &	& &\\
	    Distance (kpc) & 2.1$\pm$0.5& 1 &0.66$\pm$0.10& 2&1.19$\pm$1.15& 3\\
	    Reddening (E(B-V)) & 0.10$\pm$0.07 & 3 & 0.06$\pm$0.02& 4 & 0.08$\pm$0.03& 3\\
		Model atmosphere & Rauch  & & Rauch &  & Rauch&\\
		Effective temperature ($\mathrm{K}$) & 65\,000 & & 80\,000 & &100\,000 &\\
		Log g & 5.00 &  & 5.03 & &5.42 &\\
		Luminosity ($\mathrm{L_{\sun}}$)& 6009 &  & 5490 & &5250 &\\
		Initial mass ($\mathrm{M_{\sun}}$)& 0.605 & &0.605 & &0.615 &\\
		 &&&&&&\\
	 \textit{Nebular properties}& & &&	&\\
		$v_{shell}$ (km\,s$^{-1}$) & 33.8$\pm$1.0 & 5 & 13$\pm$1 & 2 & 27$\pm$1.0 & 6 \\
		Kinetic age (years)& 4500 &  & 4600  & &3080 &\\
		Shell radius (pc) & 0.152 &  & 0.065 & &0.085 &\\
		Inner radius (cm) & $7.85\times 10^{16}$&  &   $5.0\times 10^{16}$& & $6.31\times 10^{16}$ &\\
		Outer radius (cm) & $5.72\times 10^{17}$& & $2.09\times 10^{17}$   & &$2.66\times 10^{17}$ &\\
		Filling factor ($\epsilon$)\tnote{*} & 0.5(a); 0.1(b)&  & 0.9(a); 0.6(b); 0.4(c)  & &0.3(a); 0.6(b); 0.2(c)&\\
		Dust to gas ratio\tnote{**}  &0.5&&1.0&&1.5&\\
    	\hline 	
    	\\
		\textit{Abundances} &&  &   && &\textit{Solar\tnote{***}} \\
		He/H &$1.40\times 10^{-1}$& &$1.20\times 10^{-1}$   && $1.00\times 10^{-1}$&$8.50\times 10^{-2}$\\
		C/H &$7.80\times 10^{-4}$ & &$7.80\times 10^{-4}$  && $7.80\times 10^{-4}$&$2.69\times 10^{-4}$\\
		N/H & $3.60\times 10^{-5}$&  &$6.30\times 10^{-4}$  && $2.70\times 10^{-4}$&$6.76\times 10^{-5}$\\
		O/H &$1.32\times 10^{-4}$&  &$5.72\times 10^{-4}$ && $3.08\times 10^{-4}$&$4.89\times 10^{-4}$\\
		Ne/H & $1.10\times 10^{-4}$ & &$1.10\times 10^{-4}$  && $1.10\times 10^{-4}$&$8.51\times 10^{-5}$\\
		S/H & $2.00\times 10^{-6}$&  &$4.00\times 10^{-5}$ && $1.50\times 10^{-5}$&$1.32\times 10^{-5}$\\
		Ar/H & $2.70\times 10^{-6}$& &$2.70\times 10^{-6}$  && $2.70\times 10^{-6}$&$2.51\times 10^{-6}$\\
		\hline
	\end{tabular}
\begin{tablenotes}
	\item[*] Filling factors followed by (a), (b) and (c) indicates from inside to outside respectively, the best fit value at every sub-shell.
	\item[**] Dust to gas ratios relative to a typical ISM value =1.0.
	\item[***] From \citet{asplund09}
	\item \textbf{References}. (1) \citet{schonberner05}; (2) \citet{gomezmuñoz15}; (3) \citet{frew16}; (4) \citet{feibelman82}; (5) \citet{schonberner14};  (6) \citet{jacob13}
\end{tablenotes}
\end{threeparttable}
\end{table*}

\section{Summary and conclusions}
\label{sec_con}
Narrow band high resolution HST images were employed to study shocks and clumpiness effects on three MSPNe: NGC 3242, NGC 6826 and NGC 7662. Low and high ionisation ions were used for a deep analysis on shocked regions. Extended 2D diagnostic diagrams making use of (H$\alpha$/[\ion{N}{II}]) vs. (H$\alpha$/[\ion{S}{II}]) line ratios were studied along several sections on each nebulae (see Appendix A). We found that main deviations from the inside to outside ionisation tendency appeared only at regions where FLIERs or LISs are located.\\
Using a different approach originally proposed by \citet{guerrero13} we also searched for shocks in the [\ion{O}{III}]/H$\alpha$ line ratios at all regions within individual nebulae. Unfortunately, the low SNR at the outskirts of the studied PNe, together to uncertainties about the location of the studied profiles by \citet{guerrero13}, made us unable to reproduce their results.\\
At this small PNe sample, we observe that as much higher the slope of the fit to the (H$\alpha$/[\ion{N}{II}]) vs. (H$\alpha$/[\ion{S}{II}]) line ratios, higher is the CSPNe temperature. Considering that the studied nebulae exhibit some common properties such as sub-solar abundances, relatively small expansion velocities and high excitation ranges, it will be interesting as a future work, explore on the E2DD slope values using a bigger and diversified PNe sample.\\
Non shocked regions were selected from our earlier analysis to carried out Cloudy photoionisation models at the three nebulae. Every PNe was modelled dividing it in two or three sub-shells, where the filling factor was treated as a free parameter at each of them independently. H density laws, filling factors, dust to gas ratios and temperatures of the CSPNe were considered as free parameters and modified during the modelling. By means of a few input parameters extracted from literature, we were able to well reproduce the observed H$\alpha$ and [\ion{O}{III}] emissivity profiles.\\
In case of NGC 6826 a higher distance and thus central star temperature and evolutionary age was required to fit the line emissivity data. After several failed attempts to reach a good match using a lower distance, our best fit model was achieved for a distance of $2.1\,\mathrm{kpc}$, in agreement to previous estimations from \citet{schonberner05}, \citet{mendez90} and \citet{cudworth74}.
Comparing to previous studies, our model shows the highest central star temperature ($65\,000\,\mathrm{K}$). While H$\alpha$ and [\ion{O}{III}] are fitted very well throughout the nebula, the weaker lines [\ion{S}{II}] and [\ion{N}{II}] at the central region appears influenced by the light of the central star.\\
A very well representation of the observed [\ion{O}{III}] and H$\alpha$ emission profiles was reached for NGC 3242. Although final [\ion{S}{II}] and [\ion{N}{II}] lines do not fit very well the observations. This effect was also found in NGC 2438 by \citet{ottl14} and denoted there as inhomogeneity of the clumpiness and deviations from symmetry. Low ionisation species are strongly affected by cold clumps as also shown in infrared imaging of the molecular H$_2$ gas in the outskirts of the Helix nebula \citet{matsuura09}. Moreover, those lines are also 3 to 4 orders of magnitude weaker than [\ion{O}{III}] and H$\alpha$ in this nebula.\\
We found a very good agreement in the observed and modelled emissivity profiles for NGC 7662. Filling factor appears higher in the main body of the outer shell decreasing at the outer edge. N and S abundances were found a factor 1.5 higher that a typical PN values. Our best fit CSPN model outcomes in a $100\,000\,\mathrm{K}$, a luminosity of $5250\,L_{\sun}$, a surface gravity of $\log g=5.42$ and a initial mass of $0.615\,M_{\sun}$ of a central star. Density and ionisation structure appear comparable to the results obtained by \citet{guerrero04}.\\
On the other hand, a severe problem in studying PNe is always the distance scale. Very recent distance estimations from DR2, which were published during the review process of this investigation, differ significantly from those adopted here from previous studies.
More detailed investigations about these DR2 distances especially for planetary nebulae samples are still ongoing \citep{Stanghellini18,GAIA_PN18}. The GAIA photometric colour, resembling emission line contamination, already seem to give some hints on additional quality criteria \citep{GAIA_PN18}. Additionally, all our targets have throughout statistical errors twice as large as typically stars do have in this range of magnitudes \citep[0.04 mas according to][]{GAIA_DR2_ASTROMETRY}. It is described at section 6.3.1 in \citet{GAIA_DR2}, that for brighter sources the statistical error seems to be underestimating the real error and a bug was found in the software \textquotedblleft \dots which resulted in a serious underestimation of the uncertainties for the bright sources (G$<$13).\textquotedblright\, 
This was corrected by an ad-hoc patch only \citep[see appendix in][]{GAIA_DR2_ASTROMETRY}. Moreover, \citet{GAIA_DR2_PARALAXES} present caveats as well as known systematic effects, e.g. the slightly negative parallax zero point which is not corrected in the published data base.
\citet{Stanghellini18} mention that their analysis of DR2 sample PNe
is not reliable to an extend, to currently verify the up to now used distance scales for PNe. Our sample here is by far too small to derive a conclusion on the quality of these new distances and more detailed studies and comparisons are required. Thus, the derived physical parameters for the central stars and nebulae based on the DR2 distances, should to be used with care.
\section*{Acknowledgements}
All data used in this paper were based on observations made with the NASA/ESA Hubble Space Telescope, and obtained from the Hubble Legacy Archive, which is a collaboration between the Space Telescope Science Institute (STScI/NASA), the Space Telescope European Coordinating Facility (ST-ECF/ESA) and the Canadian Astronomy Data Centre (CADC/NRC/CSA). HST archival images were taken under proposal programme ID 6117 (P.I. Bruce Balick). This research has made use of the SIMBAD database, operated at CDS, Strasbourg, France. This work has been supporting by FONDO ALMA-Conicyt Programa de Astronom\'ia/PCI 31150001. We kindly thanks Martin Guerrero for helpful discussion.
\bibliographystyle{mnras}


\appendix
\onecolumn

\begin{figure}
\section{Sectional extended diagnostic diagrams}
Plotted below are the 36 E2DD extracted at different regions from each nebulae. Line ratios were calculated by sectors according to the observed main fragmentation structures. Regions displaying major deviations from the general inside to outside ionisation trend corresponds to zones where FLIERs or LISs are located.

		\centering \includegraphics[width=15.9cm]{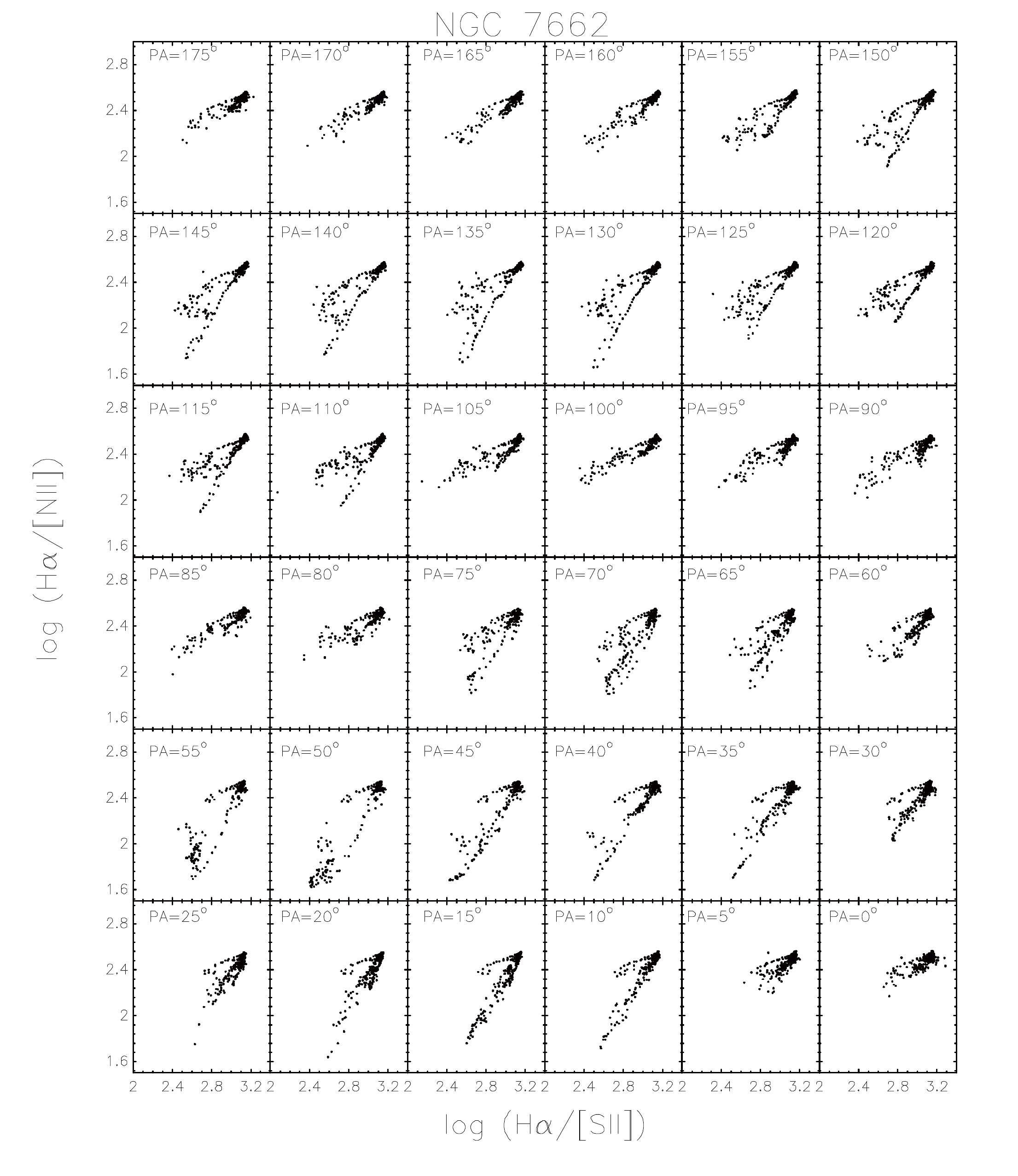}
		\caption{Sectional (H$\alpha$/[\ion{N}{II}]) vs. (H$\alpha$/[\ion{S}{II}]) line ratio diagrams for NGC 7662. Position angles are displayed at the top right on each panel. The influence of LISs is revealed along the $10\degr<PA<75\degr$ and $105\degr<PA<160\degr$.}
		\label{fig:figA1}
\end{figure}
\newpage\relax
\begin{figure}
	\centering
	\includegraphics[width=\columnwidth]{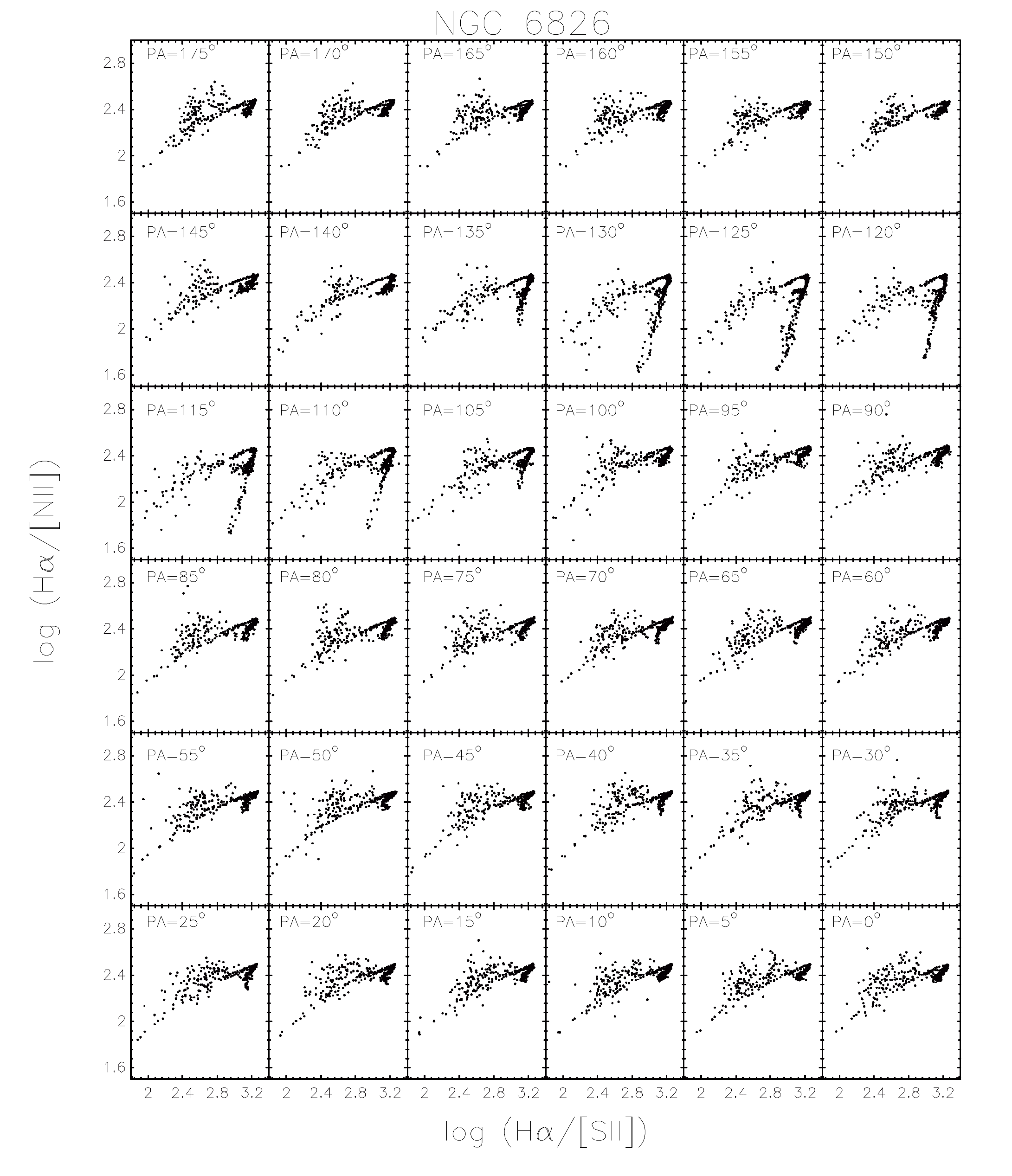}
	\caption{Same as Figure \ref{fig:figA1} but for NGC 6826. Major deviation due to FLIERs are observed at $105\degr<PA<135\degr$. }
	\label{fig:figA2}
\end{figure}

\newpage\relax
\begin{figure}
	\centering
	\includegraphics[width=\columnwidth]{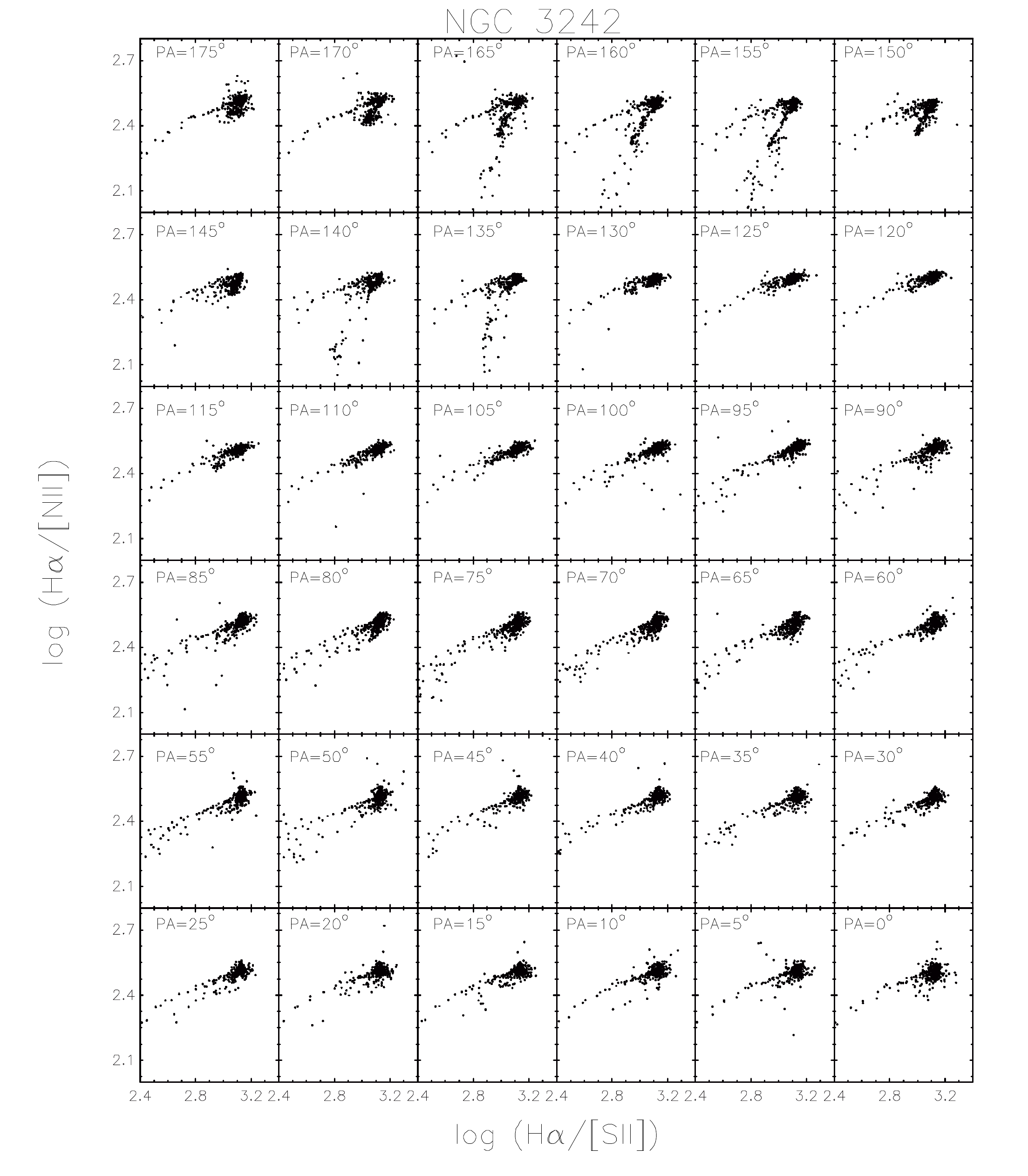}
	\caption{Same as Figure \ref{fig:figA1} but for NGC 3242. }
	\label{fig:figA3}
\end{figure}


\bsp	
\label{lastpage}
\end{document}